%% file: draft_v2.6.tex
\documentclass[twocolumn,showpacs,superscriptaddress,amsmath,amssymb,nofootinbib,aps, prd]{revtex4-2}
\usepackage{graphicx}
\usepackage{epsfig}
\usepackage{dcolumn}
\usepackage{multirow}
\usepackage{booktabs}
\usepackage{appendix}
\usepackage{makecell}
\usepackage{bm}
\usepackage{amsmath}
\usepackage{overpic}
\usepackage{comment}
\usepackage{color}
\usepackage{subfigure}
\usepackage{threeparttable}
\usepackage{natbib}
\usepackage{lineno}
\usepackage{siunitx}
\usepackage[colorlinks,linkcolor=blue,anchorcolor=blue,citecolor=blue]{hyperref}

\def\ee{e^{+}e^{-}}

\def \ee{e^+ e^-}

\def\lum{\left(\rm{pb}^{-1}\right)}

\def \gevcc{\mbox{GeV/$c^2$}}

\def \romanOne   {\uppercase\expandafter{\romannumeral1}}
\def \romanTwo   {\uppercase\expandafter{\romannumeral2}}
\def \romanThree {\uppercase\expandafter{\romannumeral3}}
\def \romanFour  {\uppercase\expandafter{\romannumeral4}}
\def \romanFive  {\uppercase\expandafter{\romannumeral5}}
\def \romanSix   {\uppercase\expandafter{\romannumeral6}}
\def \romanSeven {\uppercase\expandafter{\romannumeral7}}
\def \romanEight {\uppercase\expandafter{\romannumeral8}}
\def \romanNine {\uppercase\expandafter{\romannumeral9}}

\newcommand{\lambdacp}{\Lambda_{c}^{+}}

\def\kshort{K^0_{\mathrm{S}}}

\newcommand{\sigmode}[1]{
	\ifnum#1=1
	\lambdacp \rightarrow \Sigma^0 K^+
	\else
	\ifnum#1=2
	\lambdacp \rightarrow \Sigma^+ \kshort
	\fi
	\fi
}
\newcommand{\refmode}[1]{
	\ifnum#1=1
	\lambdacp \rightarrow \Sigma^0 \pi^+
	\else
	\ifnum#1=2
	\lambdacp \rightarrow \Sigma^+ \pi^+\pi^-
	\fi
	\fi
}
\newcommand{\sigmodefs}[1]{
	\ifnum#1=1
	\Sigma^0 K^+
	\else
	\ifnum#1=2
	\Sigma^+ \kshort
	\fi
	\fi
}
\newcommand{\refmodefs}[1]{
	\ifnum#1=1
	\Sigma^0 \pi^+
	\else
	\ifnum#1=2
	\Sigma^+ \pi^+\pi^-
	\fi
	\fi
}

\begin{document}
\title{\bf\boldmath Measurement of the Branching Fraction for the Decay  $\psi(3686) \rightarrow \phi K_{S}^{0} K_{S}^{0}$}
\input{BESIIIauthors_BAM547}
\date{\today}
\begin{abstract}
Based on  $(448.1 \pm 2.9 )\times 10^6$  $\psi(3686)$ events collected with the BESIII detector operating at the BEPCII collider, the  decay  $\psi(3686)\rightarrow \phi K_{S}^{0} K_{S}^{0}$  is observed for the first time.  Taking the interference between $\psi(3686)$ decay and  continuum production into account, the branching fraction  of this decay   is  measured  to be $\mathcal{B}(\psi(3686)\rightarrow\phi K_S^0 K_S^0 )$ = $(3.53 $ $\pm$ $0.20$ $\pm$ $0.21$)$\times 10^{-5}$, where the first uncertainty  is statistical and the second is systematic.
Combining  with the world average value for ${\mathcal B}(J/\psi\to \phi K^0_SK^0_S)$, the ratio $\mathcal{B}(\psi(3686)\rightarrow \phi K_{S}^{0} K_{S}^{0})/\mathcal{B}(J/\psi\rightarrow \phi K_{S}^{0} K_{S}^{0}) $ is determined to be   $(6.0\pm 1.6)$\%, 
which is suppressed relative to the 12\% rule.


\end{abstract}

\maketitle

\section{\boldmath Introduction}
The $J/\psi$ and $\psi(3686)$ are non-relativistic bound states of  a charm and an anti-charm quark, called  charmonium. 
Experimental measurements of the decays of charmonium states $\psi$ (which denotes both the $J/\psi$ and $\psi(3686)$)  can provide an ideal  laboratory  to study the dynamics of strong force physics, validate  models and  test various aspects of quantum chromodynamics (QCD) ~\cite{PhysRevD.99.032006,intro2}. Since the discovery of the $\psi(3686)$ in 1974~\cite{PhysRevLett.33.1453}, it has been studied for over 40 years.  However, there are still problems and puzzles that need to be understood~\cite{intro}.

Perturbative QCD predicts that both the $J/\psi $ and $\psi(3686)$ decay into light hadron final states  with a width proportional to the square of the wave
function at the origin~\cite{12rule,12rule2}.  This yields the widely-known “12\% rule”: $ \mathcal{Q}_{h}=\mathcal{B}_{\psi(3686)\rightarrow h}/\mathcal{B}_{J/\psi\rightarrow h} \simeq \mathcal{B}_{\psi(3686)\rightarrow e^{+}e^{-}}/\mathcal{B}_{J/\psi \rightarrow e^{+}e^{-}} \simeq 13.3\%$~\cite{Workman:2022ynf}, where $h$ denotes  any  exclusive  hadronic  decay  mode. Although this relation is expected to hold to a reasonably good degree for both inclusive and exclusive decays~\cite{BESIII:2020nme}, it fails severely in the case of vector-pseudoscalar meson final states, such as  $\rho \pi $ ~\cite{mark}.
With the recent experimental results on $J/\psi$ and $\psi(3686)$  two-body decays, such as the vector-tensor  channel~\cite{VT},  the pseudoscalar-pseudoscalar  channel~\cite{PP},  baryon-antibaryon mode~\cite{BB}, and  multi-body decays such as $\phi \pi^+\pi^-$, $p \bar{p} \pi^0$~\cite{multibody}, etc., extensive tests of the 12\% rule have been performed.  The ratios $\mathcal{Q}_h$ for some decay modes are suppressed, some are enhanced, while others obey the 12\% rule. More experimental results are still desirable to test the 12\% rule and further investigate charmonium decay mechanisms~\cite{intro}.

In this work, we present the first observation    and BF measurement of the decay $\psi(3686)\rightarrow \phi K_{S}^{0} K_{S}^{0}$ by analyzing  $(448.1\pm2.9)\times 10^{6}$ $\psi(3686)$ events collected with the BESIII detector in 2009 and 2012~\cite{N3686}.  In addition, the 12\% rule in the decay $\psi(3686)\to$  $\phi K_{S}^{0} K_{S}^{0}$ are tested.

\section{\boldmath BESIII Experiment and Monte Carlo Simulation}
The BESIII detector~\cite{detector} records symmetric $\ee$ collisions provided by the BEPCII~\cite{BEPCII} storage ring in the center-of-mass (CM) energy range from 2.00 to 4.95~GeV, with a peak luminosity of $1 \times 10^{33}\;\text{cm}^{-2}\text{s}^{-1}$ achieved at $\sqrt{s} = 3.77\;\text{GeV}$.

The cylindrical core of the BESIII detector covers 93\% of the full solid angle and consists of a helium-based
 multilayer drift chamber~(MDC), a plastic scintillator time-of-flight
system~(TOF), and a CsI(Tl) electromagnetic calorimeter~(EMC),
which are all enclosed in a superconducting solenoidal magnet
providing a 1.0~T magnetic field. The solenoid is supported by an
octagonal flux-return yoke with resistive plate counter muon
identification modules interleaved with steel. 
The charged-particle momentum resolution at $1~{\rm GeV}/c$ is
$0.5\%$, and the 
${\rm d}E/{\rm d}x$
resolution is $6\%$ for electrons
from Bhabha scattering. The EMC measures photon energies with a
resolution of $2.5\%$ ($5\%$) at $1$~GeV in the barrel (end cap)
region. The time resolution in the TOF barrel region is 68~ps, while
that in the end cap region is 110~ps.

Simulated data samples produced with a {\sc
geant4}-based~\cite{geant4} Monte Carlo (MC) package, which
includes the geometric description of the BESIII detector ~\cite{BesGDML, Huang_2022wuo} and the
detector response, are used to determine detection efficiencies
and to estimate backgrounds. The simulation models the beam
energy spread and initial state radiation  in the $e^+e^-$
annihilations with the generator {\sc
kkmc}~\cite{kkmc}. 

The inclusive MC sample includes the production of the $\psi(3686)$ resonance, the initial state radiation production of the $J/\psi$, and the continuum processes incorporated in {\sc
kkmc}~\cite{kkmc}. All particle decays are modelled with {\sc evtgen}~\cite{evtgen, besevtgen} using BFs
either taken from the
Particle Data Group (PDG)~\cite{Workman:2022ynf}, when available, or otherwise estimated with {\sc lundcharm}~\cite{lundcharm, Yang:2014vra}.  Final state radiation
from charged final state particles is incorporated using the {\sc
photos} package~\cite{photos}. To determine the detection efficiency for the signal process, 2 $\times$ $10^5$  signal MC samples are generated with a  modified data-driven generator  BODY3~\cite{evtgen, BODY3}, which could simulate contributions from different intermediate states in data for a given three-body final state. 
Data sets collected at center-of-mass energies ranging from 3.508 GeV to 3.773 GeV are used to estimate the contribution from continuum processes. The $\psi(3686)$ scan data sets ranging from 3.670 GeV to 3.710 GeV are used to estimate the phase angles between continuum processes and the $\psi(3686)$. The  total integrated luminosity for all data except 3.773 GeV and psi(3686) scan data  is 504.6 $\rm{pb}^{-1}$, while the  integrated  luminosity for 3.773 GeV is 2931.8 $\rm{pb}^{-1}$~\cite{lum365} and the average  integrated luminosity for $\psi(3686)$ scan data is 92.2 $ \rm{pb}^{-1}$ .

\section{\boldmath Event Selection And Data analysis }
\label{sec:selection}
To select candidate events for $\psi(3686)\rightarrow \phi K_{S}^{0} K_{S}^{0}$, the $\phi$ and $K_S^0$ mesons are reconstructed using their decays to $K^+ K^-$ and $\pi^+ \pi^-$, respectively.

Each candidate signal event is required to have at least six charged tracks. 
The charged tracks detected in the MDC are required to be within a polar angle ($\theta$) range of  $|\!\cos\theta|<0.93$, where $\theta$ is defined with respect to the $z$ axis, which is the symmetry axis of the MDC.
The charged tracks from the $\phi\rightarrow K^+K^-$  decay are required to have a distance of closest approach to  the   interaction point (IP)  less than 10 cm along  the $z$ axis ($|V_z|$) and less than 1 cm in the transverse plane ($\left| V_{xy}\right|$). 
The d$E$/d$x$ information recorded by the MDC and the time-of-flight information in the TOF are used for particle identification~(PID), and the charged tracks are assigned as  kaons  when the kaon hypothesis has a greater  likelihood, $i.e.$, $\mathcal{L}(K) $ $\textgreater$  $\mathcal{L}(\pi) $.

The charged pions used for $K^0_S$ reconstruction are required to have a  greater likelihood for the pion hypothesis, $i.e.$, $\mathcal{L}(\pi)$ $\textgreater$  $\mathcal{L}(K) $. Both pions must satisfy  $|\!\cos\theta|<0.93$ and $|V_z|$ $<$ $20$ cm and their trajectories are constrained to originate from a common vertex by applying a vertex fit~\cite{ksvertex}.  The  invariant mass of the $\pi^{+}\pi^{-}$ pair $M_{\pi^+\pi^-}$ needs to be in the range ($0.45, 0.55$) $\gevcc$. Here, $M_{\pi^+\pi^-}$ is calculated with the  pions  constrained  to  originate  at  the  decay  vertex. The $K_S^0$ candidate is then formed and the opposite direction of its momentum is constrained to point to the IP. The  decay length of the $K_S^0$ candidate is required to be greater than twice the vertex resolution.

A four-constraint (4C) kinematic fit imposing energy and momentum conservation is performed. The  helix  parameters of the charged tracks in the MC events are corrected to improve the consistency  with data. The correction factors  for $K^\pm$  are cited from Ref.~\cite{helix_cor}, while the correction factors  for $\pi^\pm$ are determined by studying the $\psi(3686)\rightarrow \pi^{+}\pi^{-} K_S^0 K_S^0$, $ K_S^0 \rightarrow \pi^+ \pi^-$ process.
The events satisfying $\chi^2_{\rm{4C}} $ $\textless$ 50 are kept for further analysis.  If there are multiple candidates in an event, the one with the smallest $\chi^2_{\rm{4C}}$ is kept for further analysis.

\begin{figure}[htbp]
 \centering
 \mbox{
 \begin{overpic}[width=\linewidth]{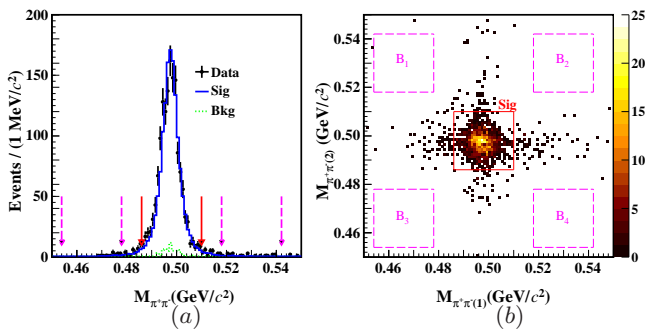}
  \put(27,-2){$(a)$}
  \put(77,-2){$(b)$}
  \end{overpic}
 }
\caption{(a) The distribution of $M_{\pi^{+}\pi^{-}}$, where the signal region is defined as $M_{\pi^{+}\pi^{-}}$ $\in$ (0.486,~0.510) $\gevcc$ and the sideband regions are defined as $M_{\pi^{+}\pi^{-}}$ $\in$ (0.454,~0.478) or (0.518,~0.542) $\gevcc$, respectively. The black points with error bars are data. The blue  solid curve is the signal MC. The green dashed curve is the inclusive MC. The pair of red arrows shows the signal region, and the two pairs of pink dashed arrows show the sideband region.
(b) The 2D distribution of $M_{\pi^{+}\pi^{-}(1)} $ vs. $ M_{\pi^{+}\pi^{-}(2)}$ (the subscripts 1 and 2 indicate the two $\pi^+\pi^-$ combinations, respectively) in the signal and sideband  regions. The red solid rectangle shows the 2D signal region where both $K_S^0$ candidates are required to be in the range (0.486,~0.510) $\gevcc$, and the pink dashed rectangles show the 2D sideband regions where both $K_S^0$ candidates are required to be in the range of (0.454,~0.478) or (0.518,~0.542) $\gevcc$.}
\label{fig:2Ddistribution}
\end{figure}

Analysis of the $\psi(3686)$ inclusive MC sample with an event type examination tool, TopoAna~\cite{topo}, indicates that the main background events come from $\psi(3686)\rightarrow  \pi^+ \pi^- J/\psi$ with $J/\psi \rightarrow \phi\pi^+\pi^-$, $\phi\rightarrow K^+K^-$. 
The background events, however, do not contain a $K_S^0K_S^0$ pair and can thus be described by $K_S^0K_S^0$ sideband events.
The one-dimensional distribution of $M_{\pi^{+}\pi^{-}}$ for the $K_S^0$ candidates in the signal and sideband regions is shown in Fig.~\ref{fig:2Ddistribution}(a). 
The  two-dimensional (2D) $M_{\pi^{+}\pi^{-}}$ distribution for the two $K_S^0$ candidates is shown in Fig.~\ref{fig:2Ddistribution}(b), where the signal region in the red solid rectangle indicates that both $K_S^0$ candidates are required to  satisfy $M_{\pi^{+}\pi^{-}}$ $\in$ (0.486,~0.510)~$\gevcc$~(marked as Sig).
The size of the signal region corresponds to three times the resolution around the known $K_S^0$ mass~\cite{Workman:2022ynf}. The 2D sideband region is defined as $M_{\pi^{+}\pi^{-}}$ $\in$  (0.454,~0.478)~$\gevcc$ or (0.518,~0.542) ~$\gevcc$ (marked as $B_i$ with $i=1,2,3,4$), where both $K_S^0$ candidates lie in the sideband region. 

\begin{figure}[htbp]
 \centering
 \mbox{
   \begin{overpic}[width=0.7\linewidth]{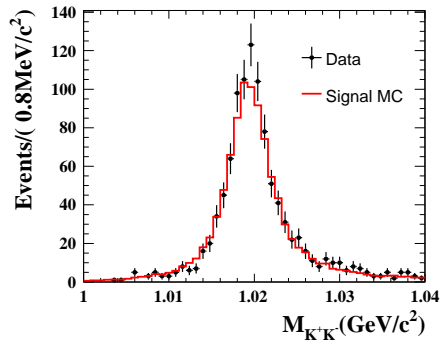}
  \end{overpic}
 }
\caption{The $M_{K^+ K^-}$ distribution for the accepted candidates in the 2D $K_S^0$  signal region for the $\psi(3686)$ data. The black points with error bars are data. The red solid curve is the signal MC.
}
\label{fig:KK_distribution}
\end{figure}

Figure~\ref{fig:KK_distribution} shows the  distributions of $M_{K^+K^-}$ in the 2D $K_S^0$ signal region. 
The number of net $\psi(3686)\rightarrow \phi K_S^0 K_S^0$ candidate
events is given by $N^{\psi(3686)}_{\rm{net}} $ = $N^{\psi(3686)}_{\rm{sig}}$ - $N^{\psi(3686)}_{\rm{bkg}}$, where  $N^{\psi(3686)}_{\rm{sig}}$ is determined by counting the events left in the 2D $K_S^0$ signal region and  $N^{\psi(3686)}_{\rm{bkg}}$ is estimated with the mean value of the background events in the sideband regions defined in Fig.~\ref{fig:2Ddistribution}(b), $i.e$., $N^{\psi(3686)}_{\rm{bkg}}$ = $\sum B_i/4$ (~with $i$ = 1, 2,~3,~4). For  $N^{\psi(3686)}_{\rm{sig}}$ and $N^{\psi(3686)}_{\rm{bkg}}$, the values are $1023\pm32$ and 0, respectively.

  \begin{figure*}[htbp!]
 \centering
 \mbox{
  \begin{overpic}[width=0.7\textwidth,
  height=0.32\textwidth]{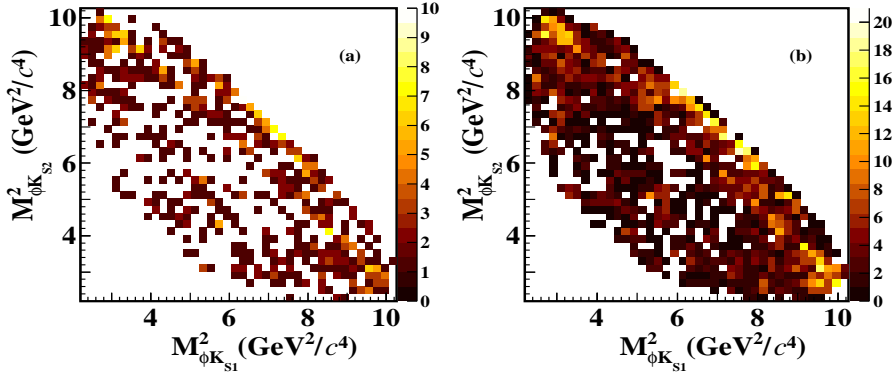}
  \end{overpic}
 }
\caption{Dalitz plots of  $M^2_{\phi K_{S1}^0}$ vs. $M^2_{\phi K_{S2}^0}$  for the accepted candidates in (a) data and (b) BODY3 signal MC events. }
\label{int:3body-data}
\end{figure*}

  \begin{figure*}[htbp!]
 \centering
 \mbox{
  \begin{overpic}[width=0.9\textwidth,
  height=0.3\textwidth]{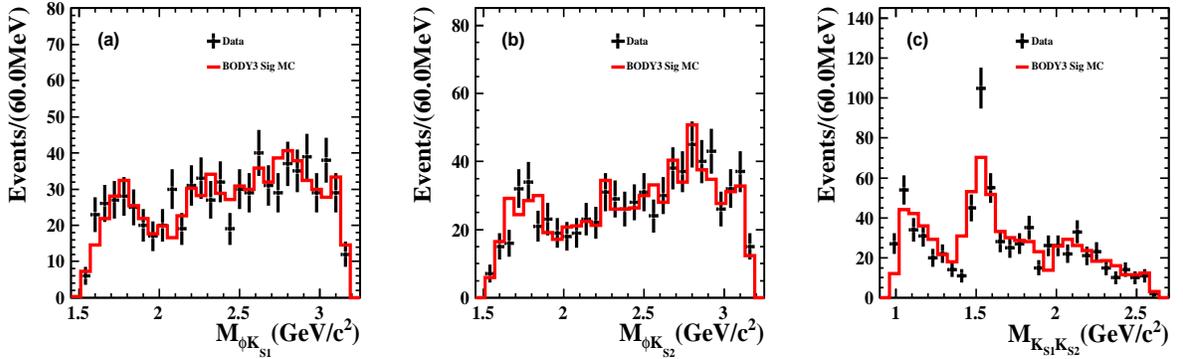}
  \end{overpic}
 }
\caption{Invariant mass distributions  for (a) $M(\phi K_{S1}^{0})$ (b) $M(\phi K_{S2}^{0})$ (c) $M(K_{S}^{0}K_{S}^{0})$. The points with error bars are data, the red histograms are the BODY3 signal MC.}
\label{fig:3body-compare}
\end{figure*}

To investigate  the  QED background from the continuum processes~\cite{N3686}, the same selection criteria and sideband
definition are applied on the off-resonance data samples.
The obtained values of signal yields $N_{\rm{net}}$  from the continuum data samples are listed in Table~\ref{tab:summary_continuum}.
A scale factor $\mathrm{f}_c$  
is considered to  account for the energy dependence of the cross section

\begin{equation}
    \mathrm{f}_{c}=\frac{\mathcal{L}_{\psi(3686)}}{\mathcal{L}_{\rm{cont.}}} \times \frac{s_{{\rm{cont.}}}^{n}}{s_{\rm{\psi(3686)}}^{ n}},
\end{equation}
 where $\mathcal{L}_{\psi(3686)}$~\cite{N3686} and $\mathcal{L}_{\rm{cont.}}$ are the integrated luminosities for the  $\psi(3686)$ and  continuum data samples,  and $s_{\psi(3686)}$ and $s_{\rm{cont.}}$ are the squares of the corresponding CM
 energies.  We use $n=1$ in the nominal result, which corresponds to a $1/s$ dependence for the continuum cross section. The impact of this assumption for $n$ will be considered as a source of systematic uncertainty.  The calculated results for $N_{\mathrm{QED}}$ estimated from different data sets are listed in the last column of Table~\ref{tab:summary_continuum}. To combine  these different $N_{\mathrm{QED}}$ into an average combined result $\overline{N}_{\mathrm{QED}}$, we use a weighted average method where the weights for each $N_{\mathrm{QED}}$ are proportional to the inverse square of the corresponding uncertainties. The $\overline{N}_{\mathrm{QED}}$ is determined to be $108\pm 5$.

Figure~\ref{int:3body-data}  shows the Dalitz plots for the accepted candidates in  data  and  BODY3  signal  MC samples.   The one-dimensional projections of the Dalitz plots are shown in Fig.~\ref{fig:3body-compare}.
The detection efficiency for $\psi(3686)\rightarrow \phi K_S^0 K_S^0$ is determined to be ($18.50\pm 0.09)\%$, where the uncertainty  comes from the MC statistics. 

The interference between the $\psi(3686)$ decay and the continuum production $e^+ e^- \rightarrow \phi K_S^0 K_S^0$
is considered by  fitting to the cross sections in the vicinity of the $\psi(3686)$, followed the method in Ref~\cite{128}. The fit yields two solutions for the phase angle between the  $\psi(3686)$ and continuum processes, corresponding to  a constructive interference of $(83\pm 11)^\circ $ and a destructive interference of $-(85 \pm 9)^\circ$. The former is determined to be  the physical one by the isospin symmetry with the decay of $\psi(3686)\rightarrow \phi K^+K^-$~\cite{Workman:2022ynf}. Hence the interference contribution is determined to be $N_{\rm{inter}}$ =$228\pm 24$.
\begin{table}[htbp!]
	 \caption{The continuum background estimation for each data set, where: $E_{\rm{CM}}$ is the center-of-mass energy;  ${\mathcal L}_{\rm cont.}$ is the integrated luminosity; $N_{\rm{net}}$ is the number of signal events in the 2D $K_S^0$ signal region after subtracting the backgrounds estimated with the sideband method; $\rm{f}_c$ is the scale factor for each energy point; and $N_{\rm{QED}}$ is the QED background calculated with $N_{\rm{net}} \times \rm{f}_c$.}
		\centering
			\begin{tabular}{l c c c c }
				\hline \hline $E_{\rm{CM}}$(GeV) & ${\mathcal L}_{\rm cont.}$$\lum$  & $N_{\text {net }}$ &$\rm{f}_c$ & $N_{\rm{QED}}$\\ \hline

			$3.508$ &$183.64$  &$32 \pm 6$& 3.30  & $106 \pm 20$    \\
			 $3.510$ &$181.79$ &$28 \pm 7$&3.34 & $94 \pm 23$ \\
			 $3.539$ &$25.50$ &$7 \pm 3$& 24.17 & $169\pm 72$\\
			 $3.553$ &$42.56$   &$10 \pm 3$&14.59 &$146 \pm 44$\\
			 $3.554$ &$27.24$  &$1 \pm 1$&22.81 &$23 \pm 23$ \\
			 $3.650$ &$43.88$  &$14 \pm 4$&14.94  &$209 \pm 60$\\
			 $3.773$ &$2931.80$  &$465 \pm 22$&0.24 &$112 \pm5$ \\
				\hline \hline
			\end{tabular}
			\label{tab:summary_continuum}
		\end{table}
The BF of  $\psi(3686)\rightarrow \phi K_S^0 K_S^0$ is determined by
\begin{equation}
\begin{split}
   &\mathcal{B}_{\psi(3686)\rightarrow \phi K_{S}^{0} K_{S}^{0}} \\
 &= \frac{N_{\rm net}^{\psi(3686)}-\overline{N}_{\mathrm{QED}}-N_{\mathrm{inter}}}{N_{\psi(3686)}\cdot \varepsilon\cdot \mathcal{B}_{\phi \rightarrow K^{+}K^{-}} \cdot \mathcal{B}_{K_{S}^{0}\rightarrow \pi^{+}\pi^{-}}^{2}},
\end{split}
\label{int:Branching_ratio}
\end{equation}
where $N_{\rm net}^{\psi(3686)}-\overline{N}_{\mathrm{QED}} - N_{\rm{inter}}$ 
=$687$ $\pm$ 40  is the net  number of $\psi(3686)\rightarrow \phi K_S^0 K_S^0$, $N_{\psi(3686)}$=$(448.1\pm2.9)\times 10^{6}$  is  the  total  number  of $\psi(3686)$ events~\cite{N3686}, $\mathcal{B}_{\phi \rightarrow K^{+}K^{-}}$ and $\mathcal{B}_{K_{S}^{0}\rightarrow \pi^{+}\pi^{-}}$ are the BFs of $\phi \rightarrow K^{+}K^{-}$ and $K_{S}^{0}\rightarrow \pi^{+}\pi^{-}$ quoted from the PDG~\cite{Workman:2022ynf}, and $\varepsilon$=(18.50 $\pm$ 0.09)\% is the detection efficiency for $\psi(3686)\rightarrow \phi K_S^0 K_S^0$. 
Based on these numbers, we  can obtain $\mathcal{B}_{\psi(3686)\rightarrow \phi K_{S}^{0} K_{S}^{0}}$ = ($3.53\pm0.20$) $\times 10^{-5}$.

\section{Systematic uncertainty }
The systematic   uncertainties  are evaluated from  a   variety   of sources,  as summarized  in  Table~\ref{int:all}.

The MDC tracking and PID efficiencies   for kaons are studied using a control sample of $J/\psi\rightarrow  K_S^0 K^\pm \pi^\mp$, $K_S^0\rightarrow \pi^+ \pi^-$. The difference in the tracking or PID  efficiencies between data and MC simulation is assigned as individual systematic uncertainties, which is 1.0\% for both tracking and PID per kaon.
The PID efficiency for pions  is determined based on studies of a control sample of $J/\psi \rightarrow \pi^+ \pi^- \pi^0$. The difference in the PID efficiencies between the data and MC simulation, 1.4\%, is assigned as the corresponding systematic uncertainty for  the four pions.

The efficiency of $K_{S}^{0}$ reconstruction is estimated  using a control sample of $J/\psi\rightarrow  K^{*}(892)^{\mp}  K^{\pm} $, $K^{*}(892)^{\mp} \rightarrow K_S^0 \pi^{\mp}$ ~\cite{PhysRevD.92.112008}. The uncertainty includes the tracking efficiency  for $\pi^+ \pi^-$, the requirement of $M_{\pi^+ \pi^-}$ and the requirement on the $K_S^0$ decay length.  The difference in the $K^0_S$ reconstruction efficiencies  between the data and MC simulation, 1.2\% per $K_S^0$, is taken as the systematic uncertainty, which is assigned to be 2.3\%.

The number of $\psi(3686)$  events is determined from an analysis of inclusive hadronic $\psi(3686)$  decays.  The  uncertainty of the number of $\psi(3686)$ events, 0.7\%~\cite{N3686}, is taken as the systematic uncertainty.
The systematic uncertainty of the BODY3  MC  model comes from the range and the bin divisions of  the input Dalitz plot. The uncertainty is estimated by changing the bin size by 20\% and taking alternative ranges at the same time. The maximum change of efficiency is taken as  the uncertainty, which is 3.6\%.

The systematic uncertainty of the 4C kinematic fit is determined by varying the helix parameters by $\pm 1$ standard deviation. The maximum change of signal efficiency, 1.3\%, is taken as the corresponding systematic uncertainty. The systematic uncertainties  due to the scale factor $\rm{f}_{c}$ for the continuum background  originate from the uncertainty of luminosity and the uncertainty of  energy dependence for the QED cross section. The systematic uncertainty originating from the luminosity is estimated by recalculating the BF  after changing the  luminosity obtained by different processes. The maximum difference of the BF, 0.1\%, is taken as  the uncertainty. The systematic uncertainty due to the energy dependence relationship 
is estimated by comparing the difference between $1/s$ and $1/s^{3}$ assumptions.
The change of the re-measured BF, 0.4\%, is assigned as the corresponding systematic uncertainty.

The uncertainty due to the limited MC statistics is considered as one source of systematic uncertainty. It is evaluated to be 0.5\%. The uncertainties of  the quoted BFs of $\mathcal{B}(K_{S}^{0}\rightarrow \pi^{+}\pi^{-})$ and $\mathcal{B}(\phi\rightarrow K^{+}K^{-})$  are assigned  to be 0.2\% and 1.0\%, respectively.

The uncertainty from  the interference between $\psi(3686)$ decay and continuum production  is determined by varying the  measured phase angle $\varphi$ by $\pm 1$ standard deviation. The maximum change of the  BF, 2.1\%  is taken as the corresponding systematic uncertainty.


Assuming that all sources are independent, the total systematic uncertainty   is determined  to be 6.0\%  as  listed in Table~\ref{int:all}. 

 \begin{table}[!htbp]
 \caption{Relative systematic uncertainties in the branching fraction measurement.}
\label{int:all}
\centering
\begin{tabular}{l c}
\hline \hline  

Source  & Uncertainty (\%)\\ \hline

$K^\pm$ tracking & 2.0 \\

$K^\pm$ PID & 2.0  \\

$\pi^\pm$ PID & 1.4 \\

$K_{S}^{0} $ reconstruction& 2.3 \\

$N_{\psi(3686)}$ &  0.7 \\

BODY3 generator & 3.6 \\

4C kinematic fit   & 1.3 \\

$\rm f_{c}$ factor & 0.4 \\

MC statistics  & 0.5 \\

$\mathcal{B}^{\rm{PDG}}(K_{S}^{0}\rightarrow \pi^{+}\pi^{-})$& 0.2\\

$\mathcal{B}^{\rm{PDG}}(\phi\rightarrow K^{+}K^{-})$& 1.0 \\
Interference & 2.1  \\ 
  \hline
  {\makecell[l]{Total \\ }} & 6.0 \\

\hline \hline
\end{tabular}
\end{table}

\section{\boldmath Summary}
Using $(448.1\pm2.9)\times 10^{6}$ $\psi(3686)$ events accumulated by the BESIII detector, the decay $\psi(3686)\rightarrow \phi K_{S}^{0} K_{S}^{0}$ is observed for the first time.  Taking the interference between $\psi(3686)$ decay and  continuum production into account, its BF  is measured to be
$(3.53 $ $\pm$ $0.20$ $\pm$ $0.21$)$\times 10^{-5}$, where the first uncertainty is statistical, the second one is systematic. Using the world average of $\mathcal{B}(J/\psi\rightarrow \phi K_{S}^{0} K_{S}^{0})=(5.9\pm 1.5)\times 10^{-4}$ ~\cite{Workman:2022ynf}, the ratio between the two BFs  is determined to be 
$
     \mathcal{Q}_{\phi K_S^0 K_S^0}  =(6.0 \pm 1.6)$ \%, which is suppressed relative to the 12\% rule. 
The  $2.7 \times 10^9$ $\psi(3686)$ events recently collected 
by BESIII~\cite{BESIII:2020nme} offer an opportunity to improve the precision of $\mathcal{Q}_h$ and will lead to a better understanding of the phenomenon.

\acknowledgments
The BESIII collaboration thanks the staff of BEPCII and the IHEP computing center for their strong support. This work is supported in part by National Key R\&D Program of China under Contracts Nos. 2020YFA0406400, 2020YFA0406300; National Natural Science Foundation of China (NSFC) under Contracts Nos. 11635010, 11735014, 11835012, 11935015, 11935016, 11935018, 11961141012, 12022510, 12025502, 12035009, 12035013, 12061131003, 12192260, 12192261, 12192262, 12192263, 12192264, 12192265; the Chinese Academy of Sciences (CAS) Large-Scale Scientific Facility Program; the CAS Center for Excellence in Particle Physics (CCEPP); Joint Large-Scale Scientific Facility Funds of the NSFC and CAS under Contract No. U1832207; CAS Key Research Program of Frontier Sciences under Contracts Nos. QYZDJ-SSW-SLH003, QYZDJ-SSW-SLH040; 100 Talents Program of CAS; The Institute of Nuclear and Particle Physics (INPAC) and Shanghai Key Laboratory for Particle Physics and Cosmology; ERC under Contract No. 758462; European Union's Horizon 2020 research and innovation programme under Marie Sklodowska-Curie grant agreement under Contract No. 894790; German Research Foundation DFG under Contracts Nos. 443159800, 455635585, Collaborative Research Center CRC 1044, FOR5327, GRK 2149; Istituto Nazionale di Fisica Nucleare, Italy; Ministry of Development of Turkey under Contract No. DPT2006K-120470; National Research Foundation of Korea under Contract No. NRF-2022R1A2C1092335; National Science and Technology fund; National Science Research and Innovation Fund (NSRF) via the Program Management Unit for Human Resources \& Institutional Development, Research and Innovation under Contract No. B16F640076; Polish National Science Centre under Contract No. 2019/35/O/ST2/02907; Suranaree University of Technology (SUT), Thailand Science Research and Innovation (TSRI), and National Science Research and Innovation Fund (NSRF) under Contract No. 160355; The Royal Society, UK under Contract No. DH160214; The Swedish Research Council; U. S. Department of Energy under Contract No. DE-FG02-05ER41374

\bibliographystyle{apsrev4-2}
\bibliography{mybib}
\end{document}

%% file: BESIIIauthors_BAM547.tex
\author{\small
M.~Ablikim$^{1}$, M.~N.~Achasov$^{13,b}$, P.~Adlarson$^{73}$, R.~Aliberti$^{34}$, A.~Amoroso$^{72A,72C}$, M.~R.~An$^{38}$, Q.~An$^{69,56}$, Y.~Bai$^{55}$, O.~Bakina$^{35}$, I.~Balossino$^{29A}$
% ...
\\
\vspace{0.2cm}
(BESIII Collaboration)\\
\vspace{0.2cm} {\it
$^{1}$ Institute of High Energy Physics, Beijing 100049, People's Republic of China\\
% ...
$^{13}$ G.I. Budker Institute of Nuclear Physics SB RAS (BINP), Novosibirsk 630090, Russia\\
% ...
$^{29}$ INFN Sezione di Ferrara, (A)INFN Sezione di Ferrara, I-44122, Ferrara, Italy; (B)University of Ferrara,  I-44122, Ferrara, Italy\\
% ...
$^{34}$ Johannes Gutenberg University of Mainz, Johann-Joachim-Becher-Weg 45, D-55099 Mainz, Germany\\
$^{35}$ Joint Institute for Nuclear Research, 141980 Dubna, Moscow region, Russia\\
% ...
$^{38}$ Liaoning Normal University, Dalian 116029, People's Republic of China\\
% ...
$^{55}$ Southeast University, Nanjing 211100, People's Republic of China\\
$^{56}$ State Key Laboratory of Particle Detection and Electronics, Beijing 100049, Hefei 230026, People's Republic of China\\
% ...
$^{69}$ University of Science and Technology of China, Hefei 230026, People's Republic of China\\
% ...
$^{72}$ University of Turin and INFN, (A)University of Turin, I-10125, Turin, Italy; (B)University of Eastern Piedmont, I-15121, Alessandria, Italy; (C)INFN, I-10125, Turin, Italy\\
$^{73}$ Uppsala University, Box 516, SE-75120 Uppsala, Sweden\\
% ...
$^{b}$ Also at the Novosibirsk State University, Novosibirsk, 630090, Russia\\
% ...
}\vspace{0.4cm}}